\documentclass[apj]{emulateapj}
\usepackage{apjfonts}
\usepackage[usenames,dvips]{color}

\begin{document}
\journalinfo{Submitted to ApJL}
\shorttitle{Variable Compact Emission in Sgr~A*}
\shortauthors{Fish et al.}
\title{1.3~\lowercase{mm} Wavelength VLBI of Sagittarius A*: Detection
  of Time-Variable Emission on Event Horizon Scales}
\author{Vincent L.\ Fish\altaffilmark{1},
        Sheperd S.\ Doeleman\altaffilmark{1}, 
        Christopher Beaudoin\altaffilmark{1},
        Ray Blundell\altaffilmark{2},
        David E.\ Bolin\altaffilmark{1,3},
        Geoffrey C.\ Bower\altaffilmark{4},
        Richard Chamberlin\altaffilmark{5},
        Robert Freund\altaffilmark{6},
        Per Friberg\altaffilmark{7},
        Mark A.\ Gurwell\altaffilmark{2},
        Mareki Honma\altaffilmark{8},
        Makoto Inoue\altaffilmark{9},
        Thomas P.\ Krichbaum\altaffilmark{10},
        James Lamb\altaffilmark{11},
        Daniel P.\ Marrone\altaffilmark{12,13},
        James M.\ Moran\altaffilmark{2},
        Tomoaki Oyama\altaffilmark{8},
        Richard Plambeck\altaffilmark{4},
        Rurik Primiani\altaffilmark{2},
        Alan E.\ E.\ Rogers\altaffilmark{1},
        Daniel L.\ Smythe\altaffilmark{1},
        Jason SooHoo\altaffilmark{1},
        Peter Strittmatter\altaffilmark{6},
        Remo P.\ J.\ Tilanus\altaffilmark{7,14},
        Michael Titus\altaffilmark{1},
        Jonathan Weintroub\altaffilmark{2},
        Melvyn Wright\altaffilmark{4},
        David Woody\altaffilmark{11},
        Ken H.\ Young\altaffilmark{2},
        Lucy M.\ Ziurys\altaffilmark{6}
}
\email{vfish@haystack.mit.edu}
\altaffiltext{1}{Massachusetts Institute of Technology, Haystack
  Observatory, Route 40, Westford, MA 01886, USA}
\altaffiltext{2}{Harvard-Smitshonian Center for Astrophysics, 60 Garden
  St., Cambridge, MA 02138, USA}
\altaffiltext{3}{University of Arizona, Tucson AZ, 85721, USA}
\altaffiltext{4}{University of California Berkeley, Dept.\ of
  Astronomy, Radio Astronomy Laboratory, 601 Campbell, Berkeley, CA
  94720-3411, USA}
\altaffiltext{5}{Caltech Submillimeter Observatory, 111 Nowelo St.,
  Hilo, HI 96720, USA}
\altaffiltext{6}{Arizona Radio Observatory, Steward Observatory,
  University of Arizona, 933 North Cherry Ave., Tucson, AZ 85721-0065,
  USA}
\altaffiltext{7}{James Clerk Maxwell Telescope, Joint Astronomy Centre,
  660 North A'ohoku Place, University Park, Hilo, HI 96720, USA}
\altaffiltext{8}{National Astronomical Observatory of Japan, Osawa
  2-21-1, Mitaka, Tokyo 181-8588, Japan}
\altaffiltext{9}{Institute of Astronomy and Astrophysics, Academia
  Sinica, P.O. Box 23-141, Taipei 10617, Taiwan}
\altaffiltext{10}{Max-Planck-Institut f\"{u}r Radioastronomie, Auf dem
  H\"{u}gel 69, D-53121 Bonn, Germany }
\altaffiltext{11}{OVRO, California Institute of Technology, 100 Leighton
  Lane, Big Pine, CA 93513-0968, USA}
\altaffiltext{12}{Kavli Institute for Cosmological Physics, University
  of Chicago, 5640 South Ellis Avenue, Chicago, IL 60637, USA}
\altaffiltext{13}{Hubble Fellow}
\altaffiltext{14}{Netherlands Organization for Scientific Research, Laan
  van Nieuw Oost-Indie 300, NL2509 AC The Hague, The Netherlands}

\begin{abstract}
Sagittarius~A*, the $\sim4\times10^6~M_\odot$ black hole candidate at
the Galactic Center, can be studied on Schwarzschild radius scales
with (sub)millimeter wavelength Very Long Baseline Interferometry
(VLBI).  We report on 1.3~mm wavelength observations of Sgr~A* using a
VLBI array consisting of the JCMT on Mauna Kea, the ARO/SMT on
Mt.~Graham in Arizona, and two telescopes of the CARMA array at Cedar
Flat in California.  Both Sgr~A* and the quasar calibrator 1924$-$292
were observed over three consecutive nights, and both sources were
clearly detected on all baselines.  For the first time, we are able to
extract 1.3~mm VLBI interferometer phase information on Sgr~A* through
measurement of closure phase on the triangle of baselines.  On the
third night of observing, the correlated flux density of Sgr~A* on all
VLBI baselines increased relative to the first two nights, providing
strong evidence for time-variable change on scales of a few
Schwarzschild radii.  These results suggest that future VLBI
observations with greater sensitivity and additional baselines will
play a valuable role in determining the structure of emission near the
event horizon of Sgr~A*.

\end{abstract}

\section{Introduction}

The case for linking Sgr~A*, the radio source at the center of the
Milky Way, with a supermassive black hole is very strong.  Mass
estimates inferred from stellar orbits, proper motion studies that
indicate Sgr~A* is nearly motionless, VLBI observations that reveal it
is ultracompact, and short-timescale variability from radio to X-rays
all point towards Sgr~A*'s association with a
$\sim4\times10^6~M_\odot$ black hole \citep[and references
  therein]{reid2009}.  At a distance of $\sim8~$kpc, the Schwarzschild
radius of this black hole subtends $R_\mathrm{Sch} \sim10~\mu$as,
making the apparent size of its event horizon the largest that we know
of.  VLBI at (sub)millimeter wavelengths is ideally suited to
observing Sgr~A* on these angular scales.  Previous 1.3~mm VLBI
detections of Sgr~A* on a Hawaii-Arizona baseline established the
existence of coherent structures on scales of a few $R_\mathrm{Sch}$
\citep{doeleman2008}.

Current 1.3~mm VLBI observations can be used to address two
fundamental questions concerning the nature of Sgr~A*.  The first is
whether the accretion flow surrounding Sgr~A* exhibits an expected
``shadow'' feature that occurs due to the strong gravitational lensing
near the black hole.  Emission from the accretion flow is
preferentially lensed onto the last photon orbit, resulting in a
relatively dim central region encircled by a brighter annulus
\citep{falcke2000}.  A second question is whether the flaring behavior
exhibited by Sgr~A* has its origins in compact structures that arise
near the black hole event horizon.  Broadband flares on timescales
ranging from minutes to hours are well-documented
\citep{marrone2008,yusefzadeh2009,doddseden2009} and imply
time-variable structures in the innermost accretion region.
If small-scale variable structures are present, 1.3~mm VLBI can
sensitively monitor the changing morphology of Sgr~A* using
non-imaging techniques with time resolutions of tens of seconds
\citep{doeleman2009,fish2009b}.

We report on new 1.3~mm VLBI observations of Sgr~A* using a
four-telescope array.  These observations confirm event horizon scale
structure within Sgr~A*, impose new constraints on accretion models
for Sgr~A*, and reveal time-dependent variability on scales of a few
$R_\mathrm{Sch}$.

\section{Observations}
\label{observations}

Sgr~A* and several calibrator sources were observed with four
telescopes at three observatories: the James Clerk Maxwell Telescope
(JCMT; henceforth also J) on Mauna Kea in Hawaii, the Arizona Radio
Observatory's Submillimeter Telescope (ARO/SMT; S) in Arizona, and two
telescopes of the Combined Array for Research in Millimeter-wave
Astronomy (CARMA; C and D, located $\sim60$~m apart) in California.
On Mauna Kea, the Submillimeter Array (SMA) housed the VLBI recording
system and synthesized the hydrogen maser based VLBI reference used at
the JCMT.  Masers at all sites were checked against ultra-stable
crystals; combined losses due to maser instabilities and local
oscillator decoherence are estimated to be $\lesssim5$\%.
Observations occurred over three nights: 2009 April 5--7 (days
95--97).  Sources were observed in left circular polarization in two
480~MHz bandwidths centered at 229.089 and 229.601~GHz (low and high
bands).  Data recorded at all sites was shipped to MIT Haystack
Observatory in Westford, Massachusetts for processing on the Mark4
VLBI correlator.  Once correlated, data for each scan (typically
10--15 minutes) were corrected for coherence losses due to atmospheric
turbulence and searched for detections using methods detailed in
\citet{doeleman2001, doeleman2008}.  Atmospheric coherence times
ranged from a few to $\sim20$~s, depending on weather conditions at
each telescope.

\section{Calibration}\label{calibration}
  
The VLBI correlation coefficient for each baseline was multiplied by
the geometric mean of the System Equivalent Flux Density (SEFD) of
both antennas.
The SEFD is a product of antenna gain (Jy/K) and the opacity-corrected
system temperature, which
was measured just prior to each VLBI scan using a
vane calibration technique that corrects for the atmosphere.  For the
JCMT and ARO/SMT, antenna gains were determined from observations of
planets at several points during the multiple day campaign, and the
gains were observed to be stable.  Relative gains for the two CARMA
dishes were estimated using observations taken by CARMA in
interferometric array mode before each VLBI scan, and the gains were
then set to a common flux scale using planet scans at the end of each
night.

The flux densities of all VLBI targets (Sgr~A*, 1924$-$292, M87,
3C273, 3C345, 1733$-$130, 3C279, 0854$+$201) were measured with CARMA.
For Sgr~A*, data with baselines shorter than 20~k$\lambda$ were
discarded to filter out extended emission in the Galactic center.  The
measured flux densities of all sources increased from day 95 to day 96
and from day 96 to day 97.  We attribute this systematic trend to
errors in the planet calibrations made shortly after sunrise, when
antenna focus, pointing offsets, and atmospheric coherence typically
change.  The flux density measured for the calibrator 1924$-$294,
observed over the same time and elevation ranges as Sgr~A*, was 9.95,
10.21, and 10.75~Jy on days 95, 96, and 97.  We normalized CARMA gains
to a constant flux density of 10.25~Jy on all three days.  The
resulting measured flux densities for Sgr A* are 3.03, 3.16, and
3.61~Jy on days 95, 96, and 97, respectively.  We adopt these fluxes
for all subsequent analysis.

\begin{figure*}
\resizebox{0.31\hsize}{!}{\includegraphics{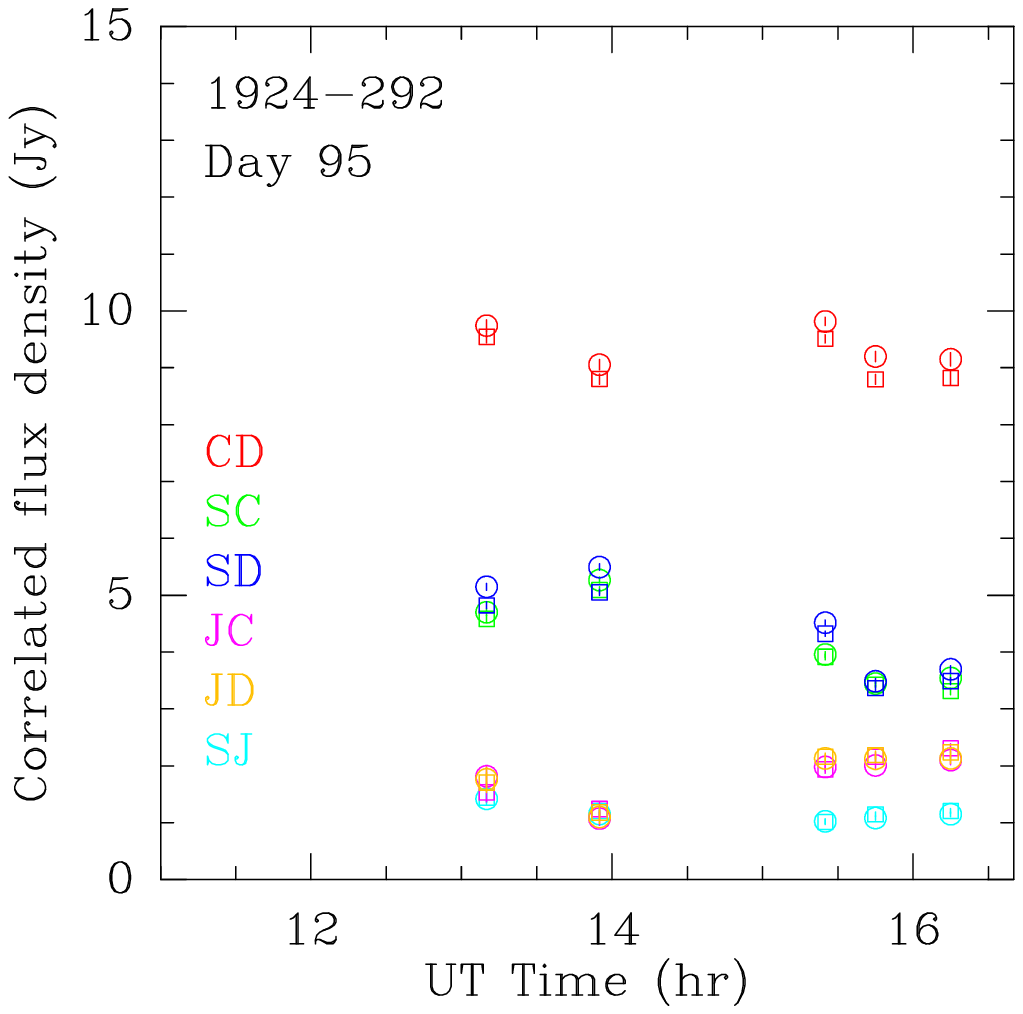}}
\resizebox{0.31\hsize}{!}{\includegraphics{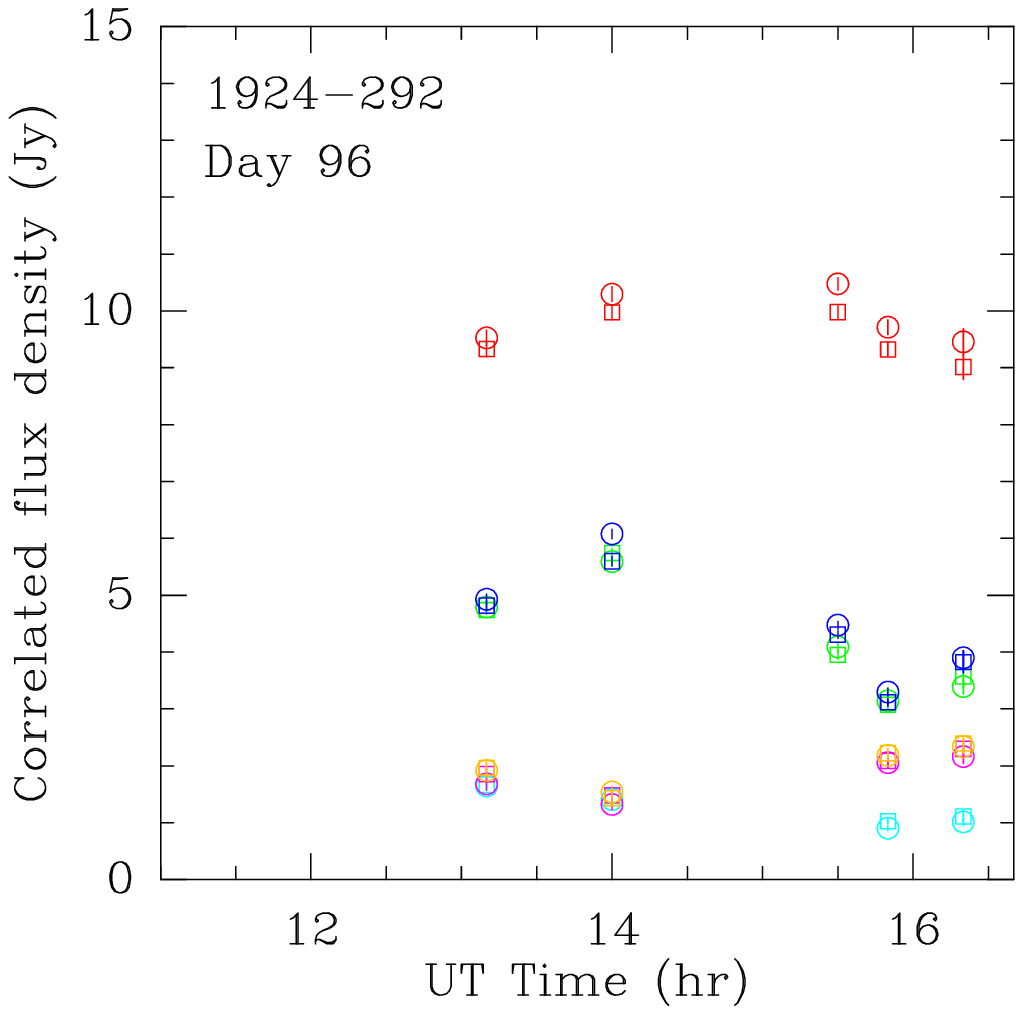}}
\resizebox{0.31\hsize}{!}{\includegraphics{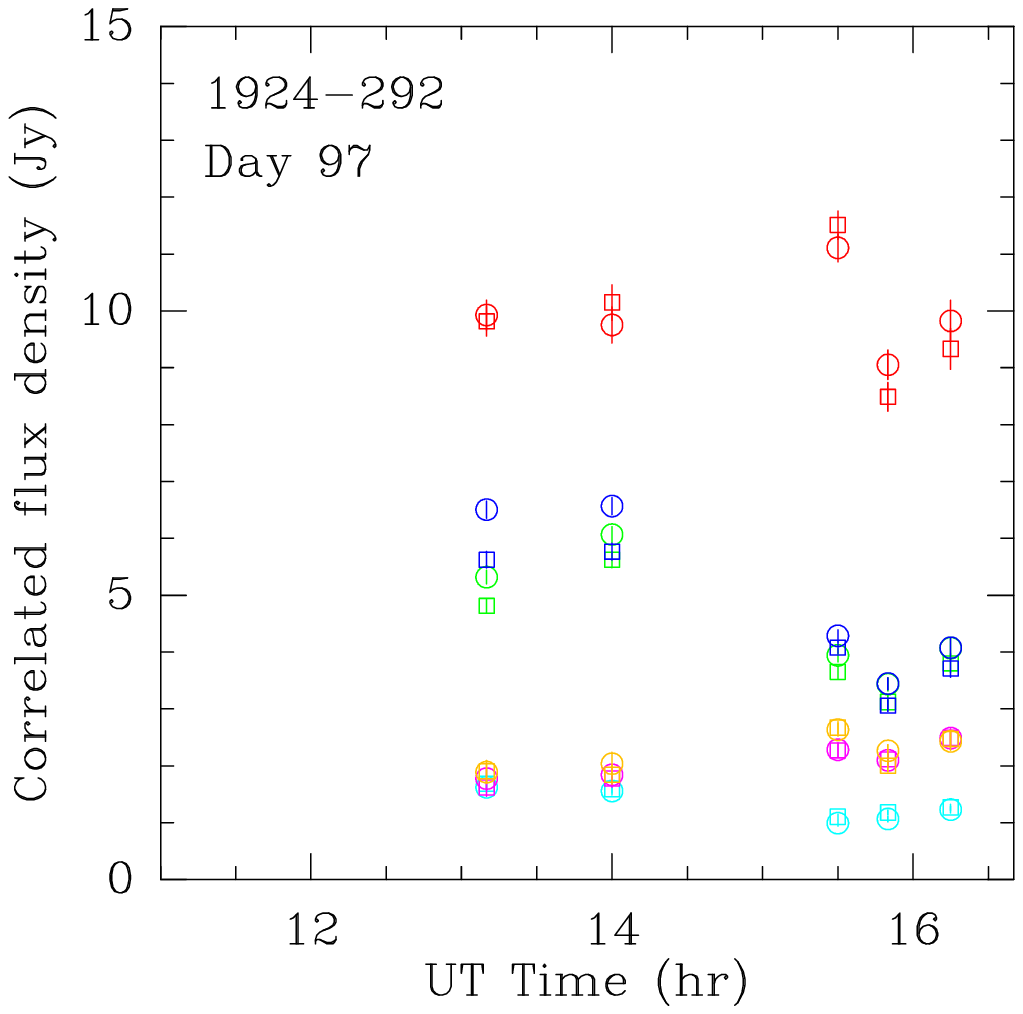}}\\
\resizebox{0.31\hsize}{!}{\includegraphics{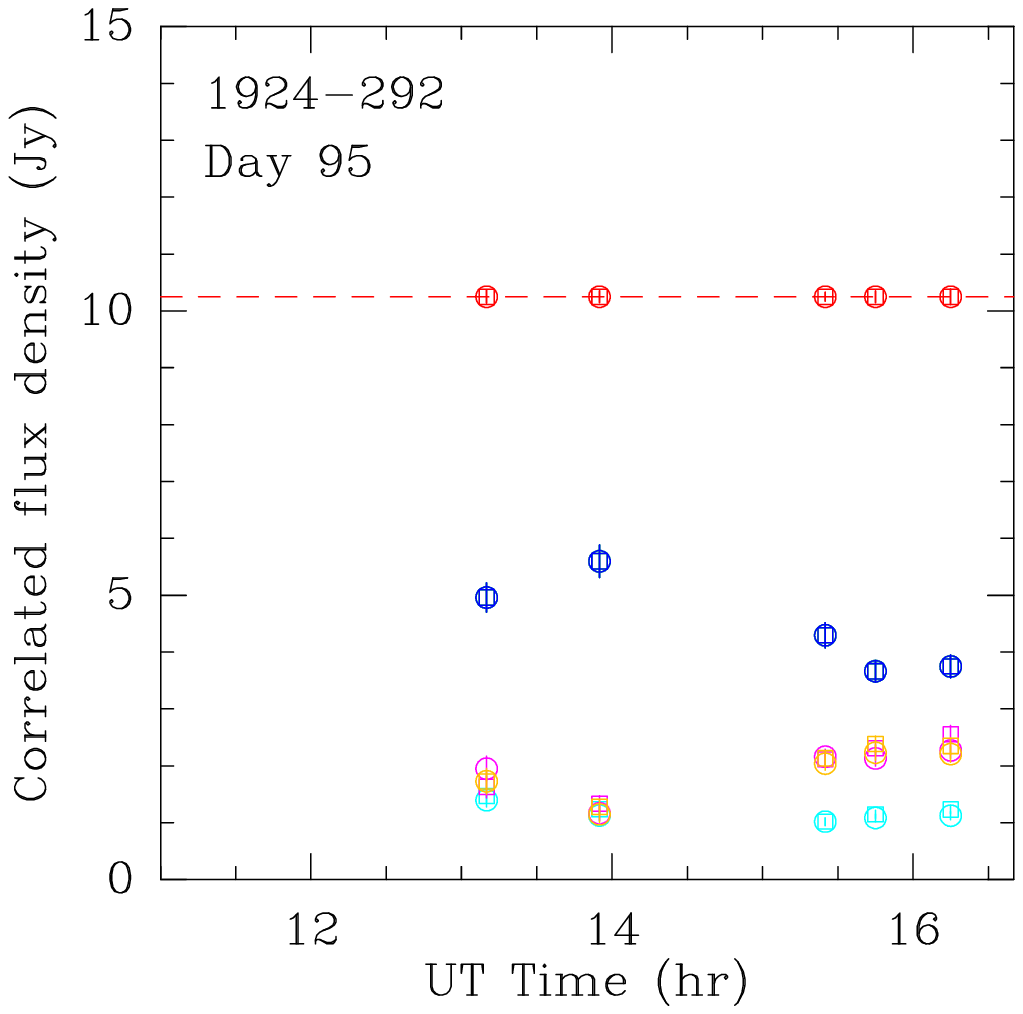}}
\resizebox{0.31\hsize}{!}{\includegraphics{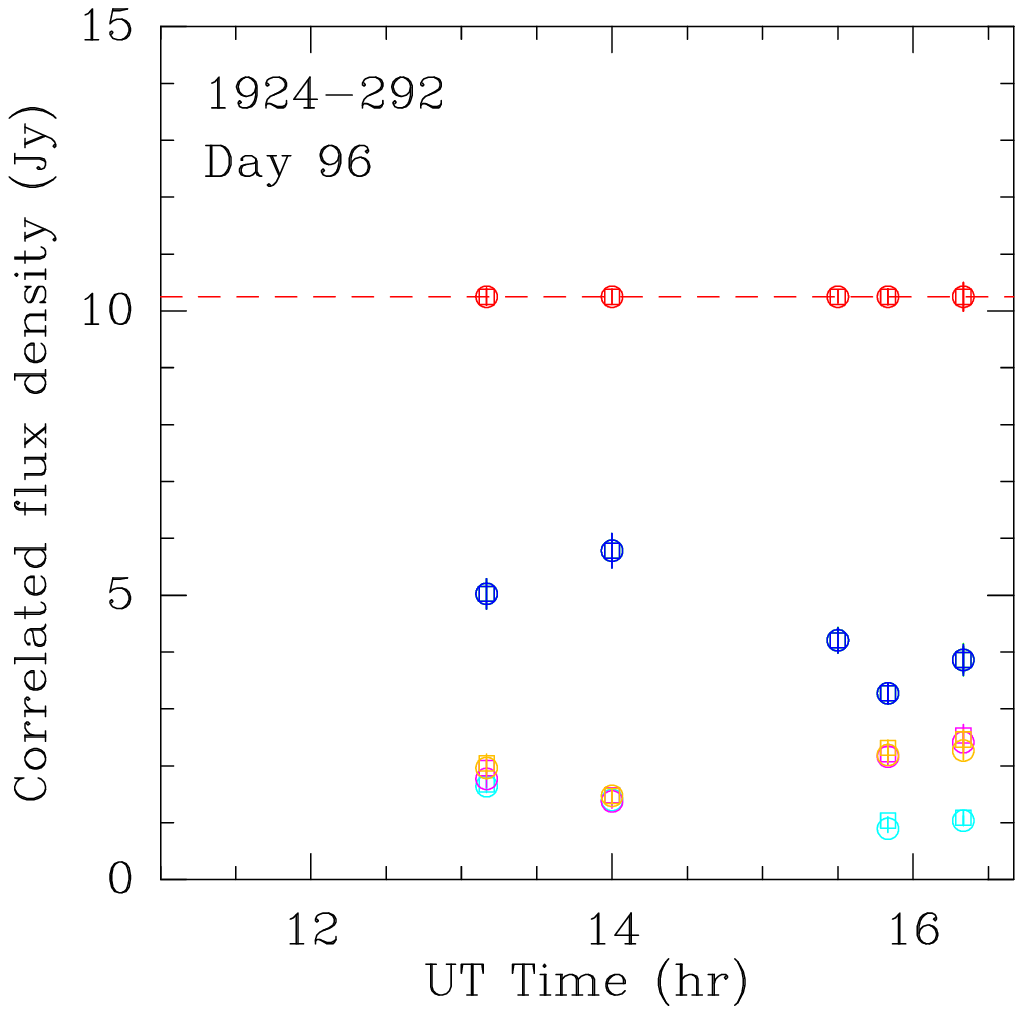}}
\resizebox{0.31\hsize}{!}{\includegraphics{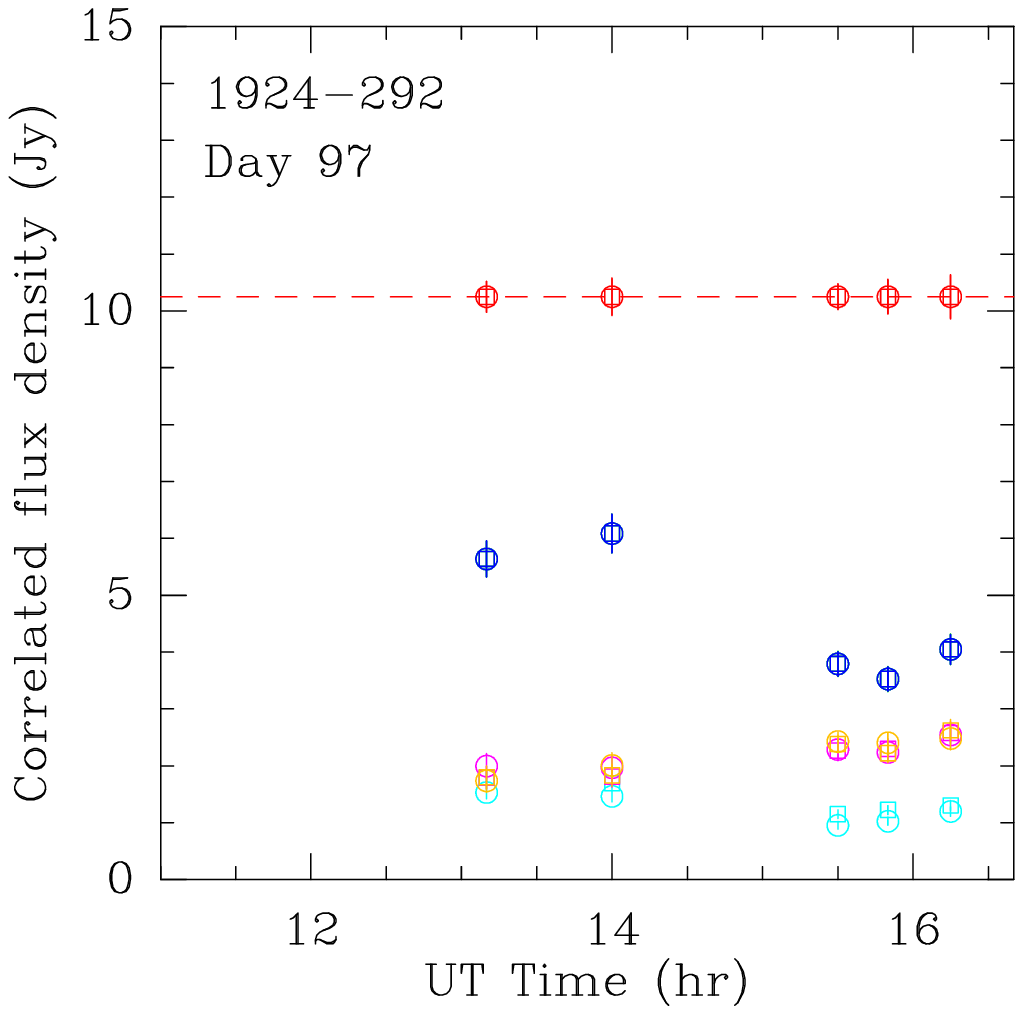}}
\caption{\emph{Top}: Observations of $1924-292$.  Detections are
  color-coded by baseline.  Squares indicate low-band data and circles
  indicate high-band data.  The day 97 a-priori calibration is
  noticeably poorer at CARMA.  \emph{Bottom}: Same, after gain
  correction.  Errorbars include a 5\% systematic component.  The
  gain-corrected data exhibit much higher day-to-day repeatability.
  The red dashed line shows the assumed CD flux (the associated flux
  scale uncertainty is $\sim5$\%, modulo uncertainties in planet
  fluxes).  Gain-corrected SC and SD data are equal by definition.
\label{fig-1921-scans}
}
\end{figure*}

As shown in the upper panels of Figures~\ref{fig-1921-scans} and
\ref{fig-sgra-scans}, there are still noticeable variations in the
correlated flux densities even after renormalizing the day-to-day flux
scales.  These residual calibration errors and amplitude variation can
be corrected for by making three simplifying assumptions that allow us
to use standard ``self-calibration" techniques.
First, the flux densities of detections in the low and high bands,
which differ by only 0.2\% in frequency, are assumed to be equal in
each scan.  Second, flux densities on the SC and SD baselines are
assumed to be equal.  While one could in principle require that JC and
JD flux densities be equal as well, the signal-to-noise ratio (SNR) is
generally much lower on the JC and JD baselines than on the shorter
VLBI baselines (SC and SD), since both 1924$-$292 and Sgr~A* are more
resolved on longer baselines.  Third, CARMA antenna gains are adjusted
to make the correlated flux density on the CD baseline (with a fringe
spacing measured in arcseconds) equal to the total flux density
measured each night by CARMA.  This final constraint enforces a
constant source flux density over the duration of each night of
observation.  While some of the observed variation in Sgr~A* over the
course of a night may be due to intrinsic variability, the 1924$-$292
data exhibit similar scatter, suggesting that calibration errors may
dominate over source variability.  Combined, these assumptions result
in a closed-form solution for gain-correction coefficients for
telescopes C, D, and S in each band.  Henceforth, we will use the term
``gain-corrected'' to refer to flux densities that have been
multiplied by these gain-correction coefficients.  We note that if the
total flux density (CD) is varied, the SJ flux densities are unchanged
while other flux densities vary as the square root of the factor.

The quasar 1924$-$292 was easily detected on all baselines
(Table~\ref{table-detections}).  On each scan, low-band and high-band
fluxes after a-priori calibration track each other consistently
(Figure~\ref{fig-1921-scans}).  After gain correction assuming a total
flux density of $10.25$~Jy, the data from all three days are highly
consistent with one another.  The SC and SD baselines show consistent
variation in the correlated flux density that is repeated each day.
The long-baseline detections (SJ, JC, and JD) also show day-to-day
repeatable behavior, indicating detection of stable source structure
presumably associated with a jet \citep{shen1997}.  The consistency of
these data demonstrates the validity of the gain-correction technique.
Based on the statistics of the data on scans of 1924$-$292, systematic
errors are estimated to be $\sim5$\%.

\section{Results}\label{results}

\begin{figure*}
\resizebox{0.31\hsize}{!}{\includegraphics{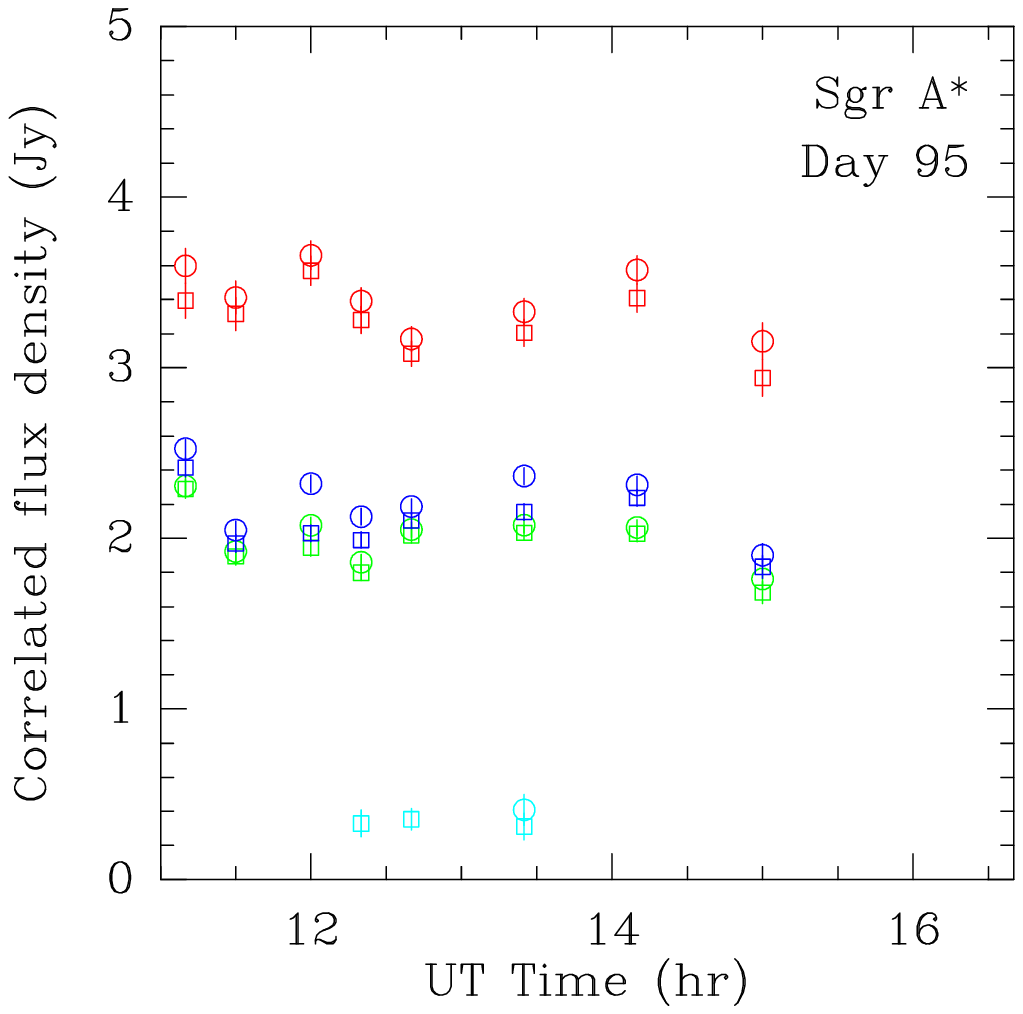}}
\resizebox{0.31\hsize}{!}{\includegraphics{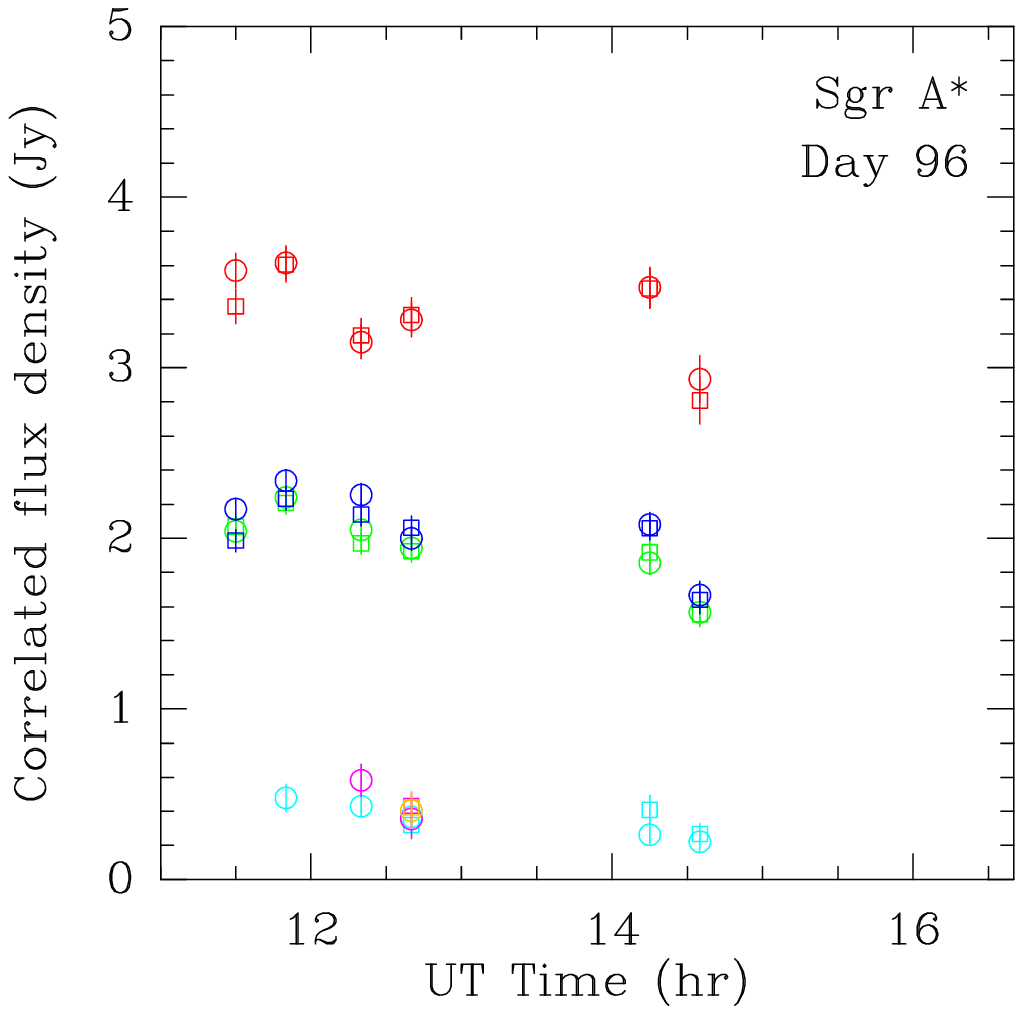}}
\resizebox{0.31\hsize}{!}{\includegraphics{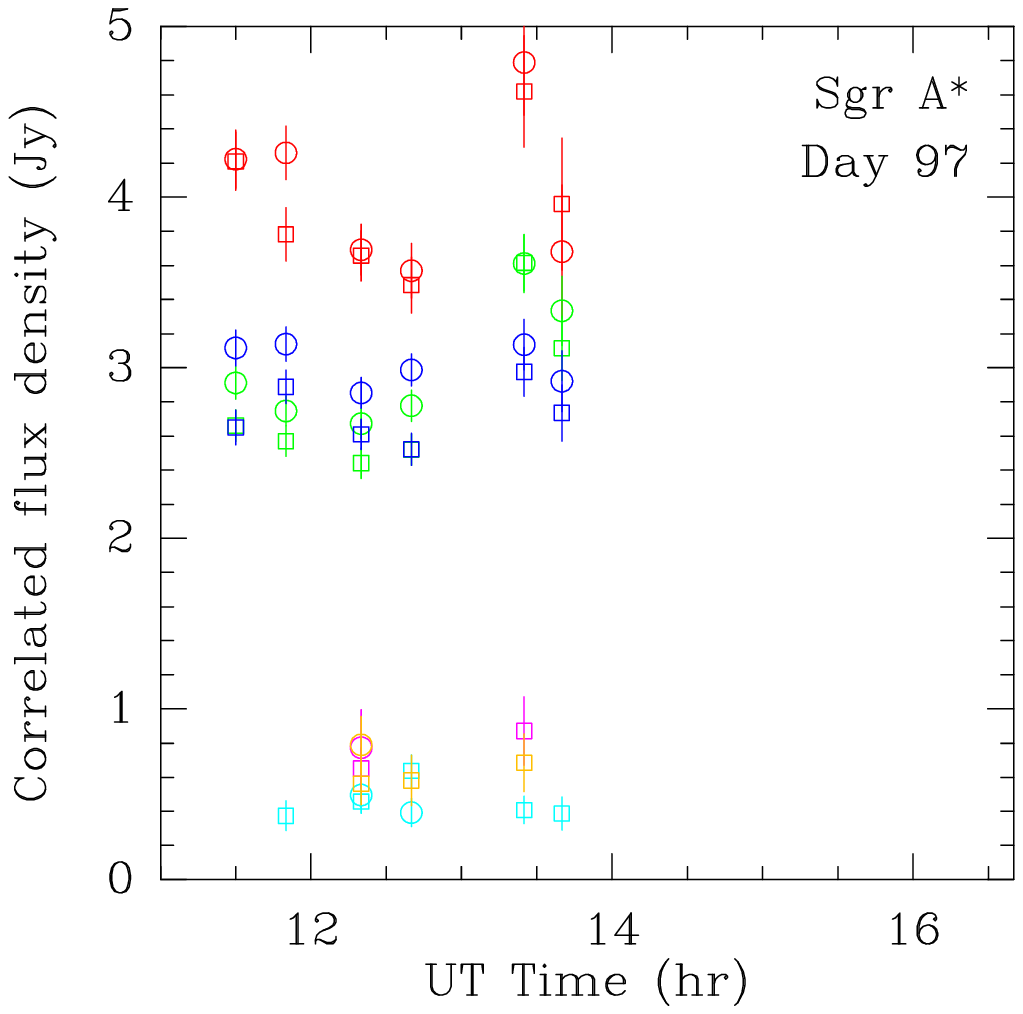}}\\
\resizebox{0.31\hsize}{!}{\includegraphics{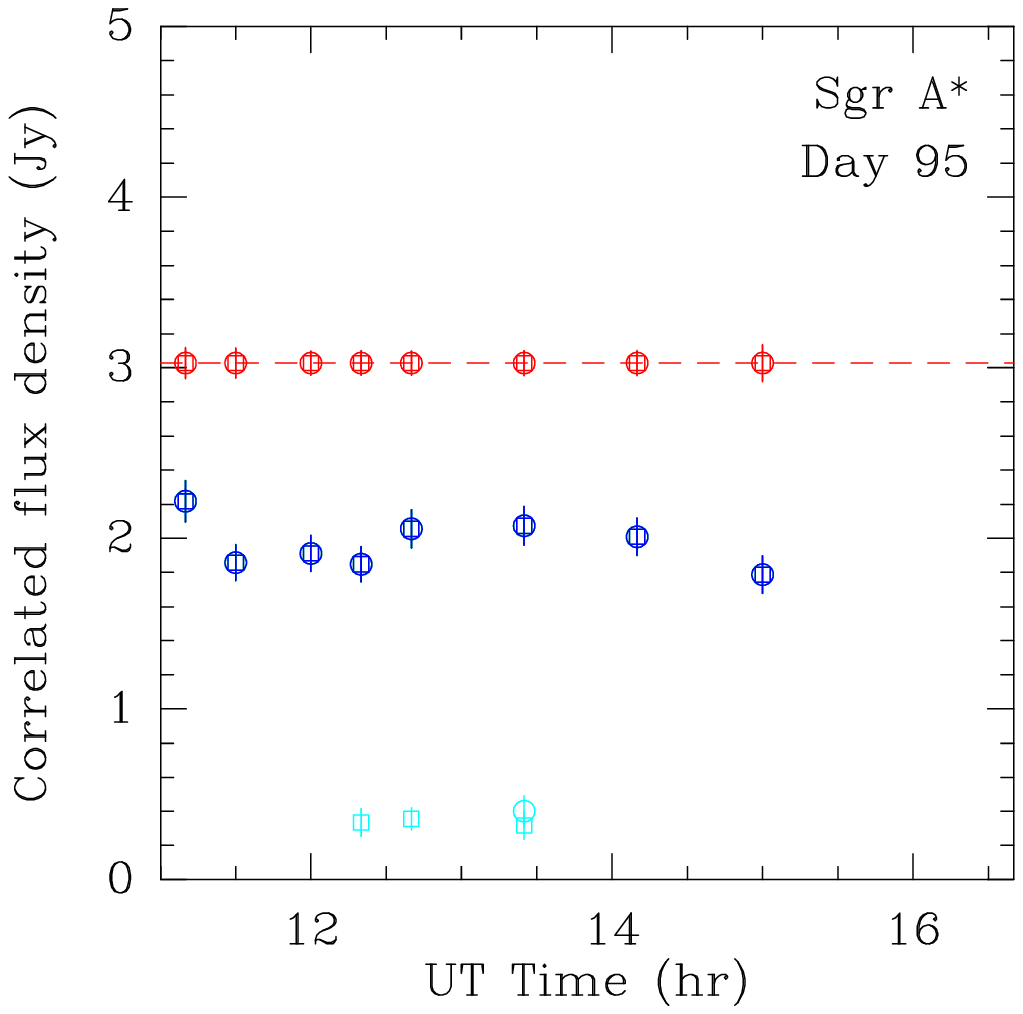}}
\resizebox{0.31\hsize}{!}{\includegraphics{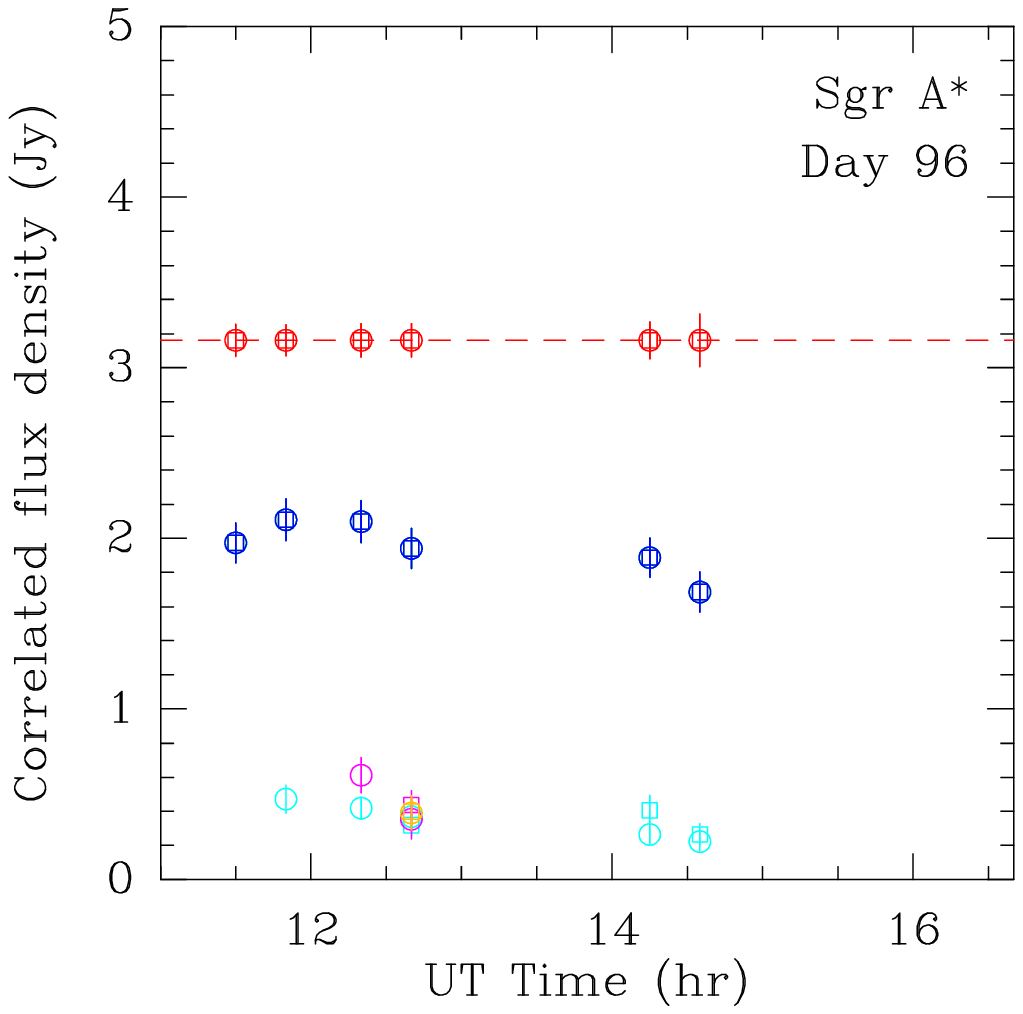}}
\resizebox{0.31\hsize}{!}{\includegraphics{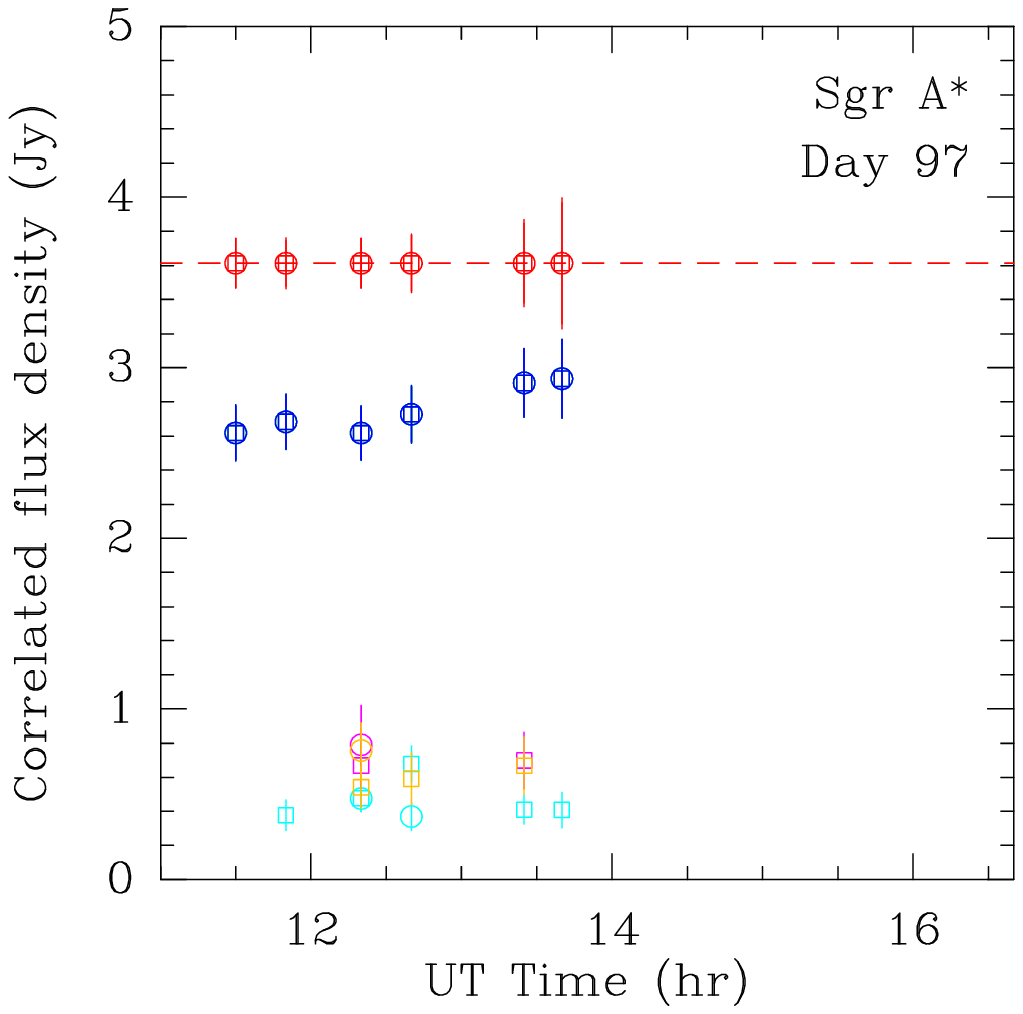}}
\caption{Same as Figure~\ref{fig-1921-scans} but for Sgr~A*.  Only
  scans for which detections are seen in both bands on the CD, SC, and
  SD baselines are shown.
\label{fig-sgra-scans}
}
\end{figure*}

\begin{figure*}
\resizebox{0.31\hsize}{!}{\includegraphics{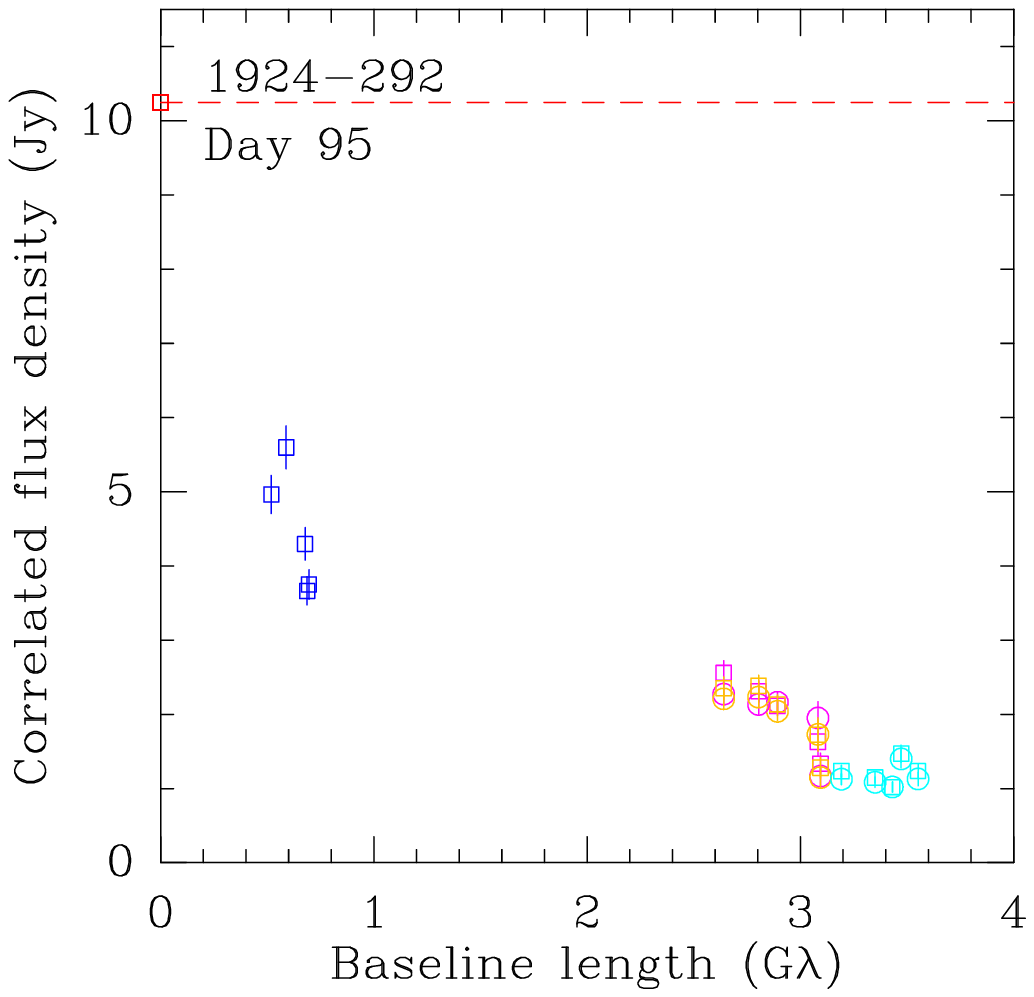}}
\resizebox{0.31\hsize}{!}{\includegraphics{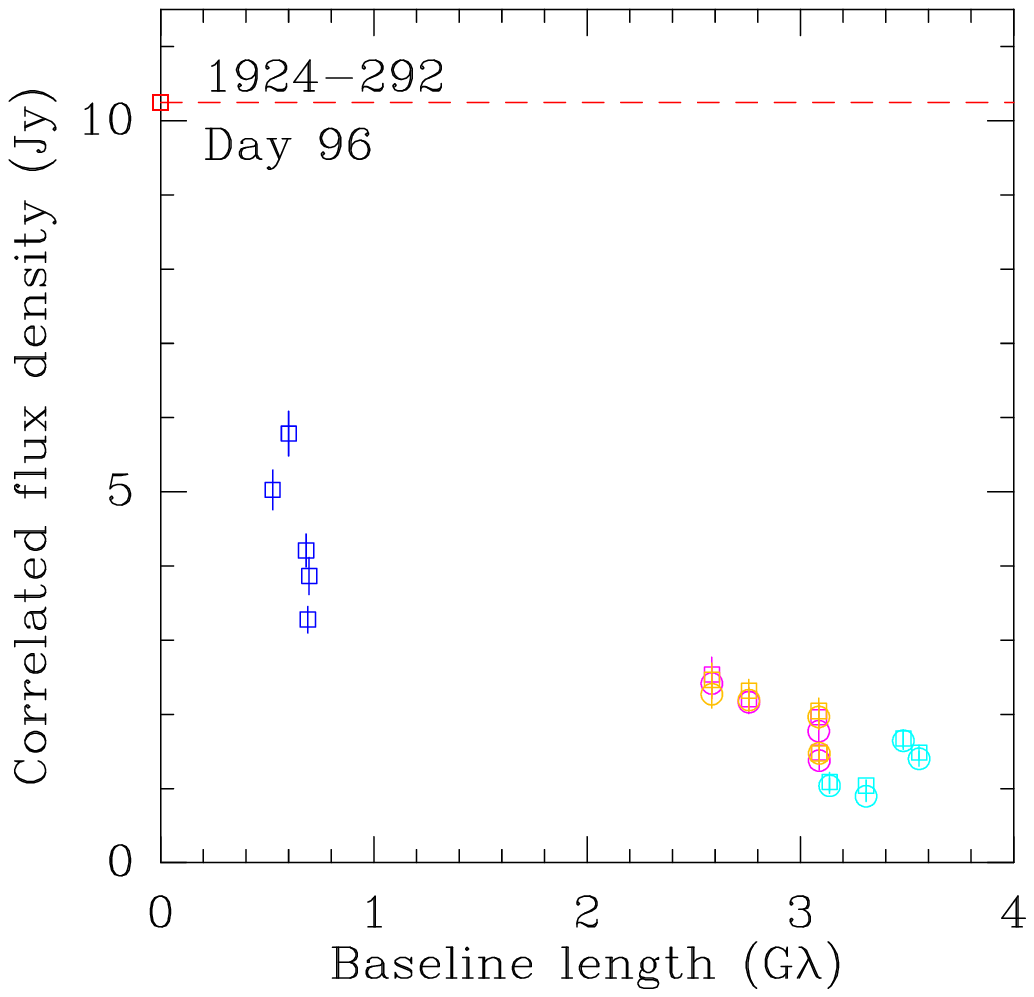}}
\resizebox{0.31\hsize}{!}{\includegraphics{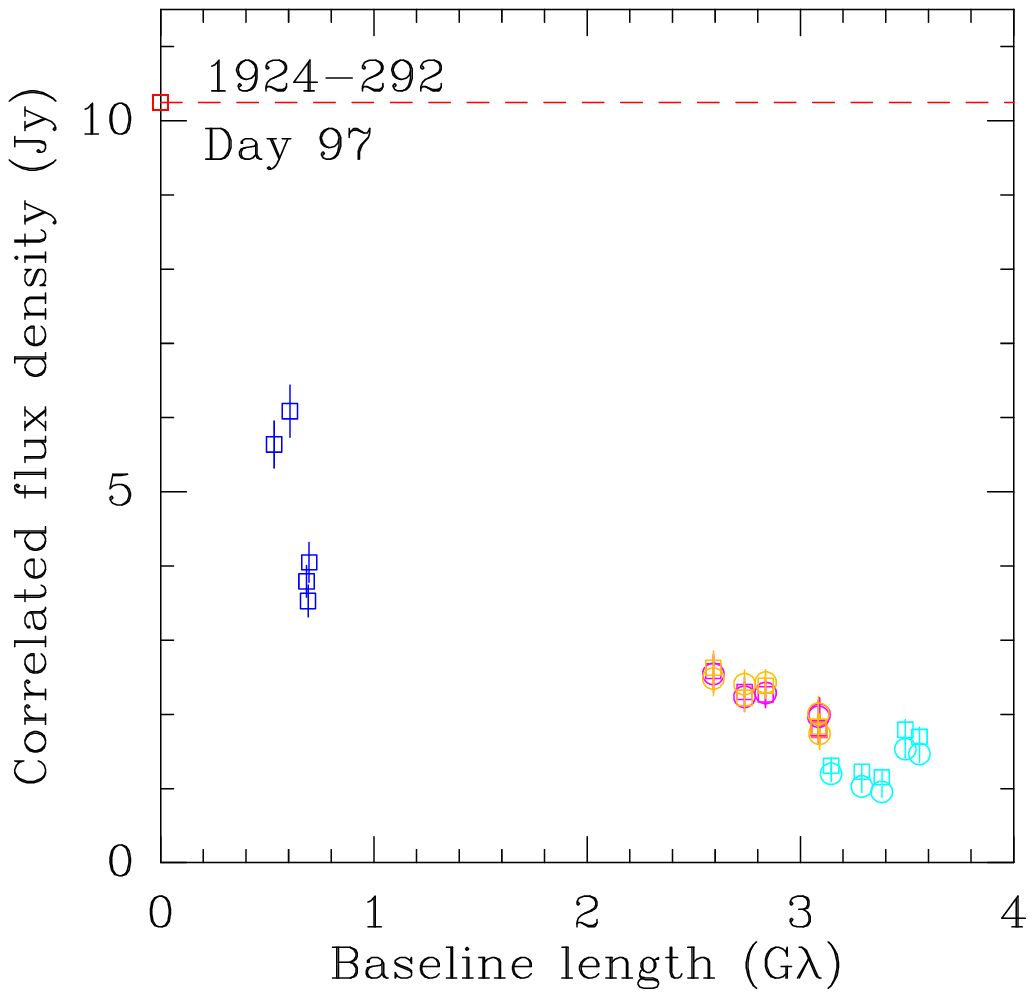}}\\
\resizebox{0.31\hsize}{!}{\includegraphics{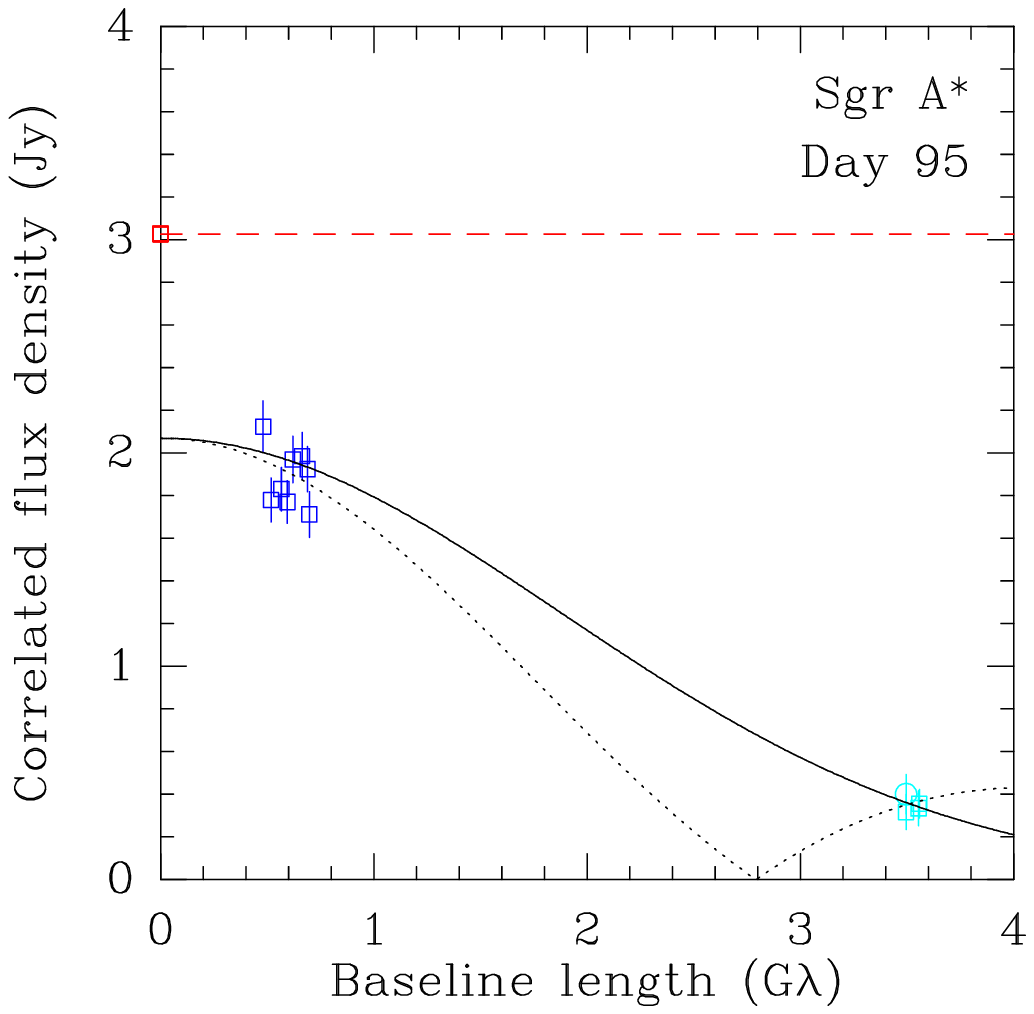}}
\resizebox{0.31\hsize}{!}{\includegraphics{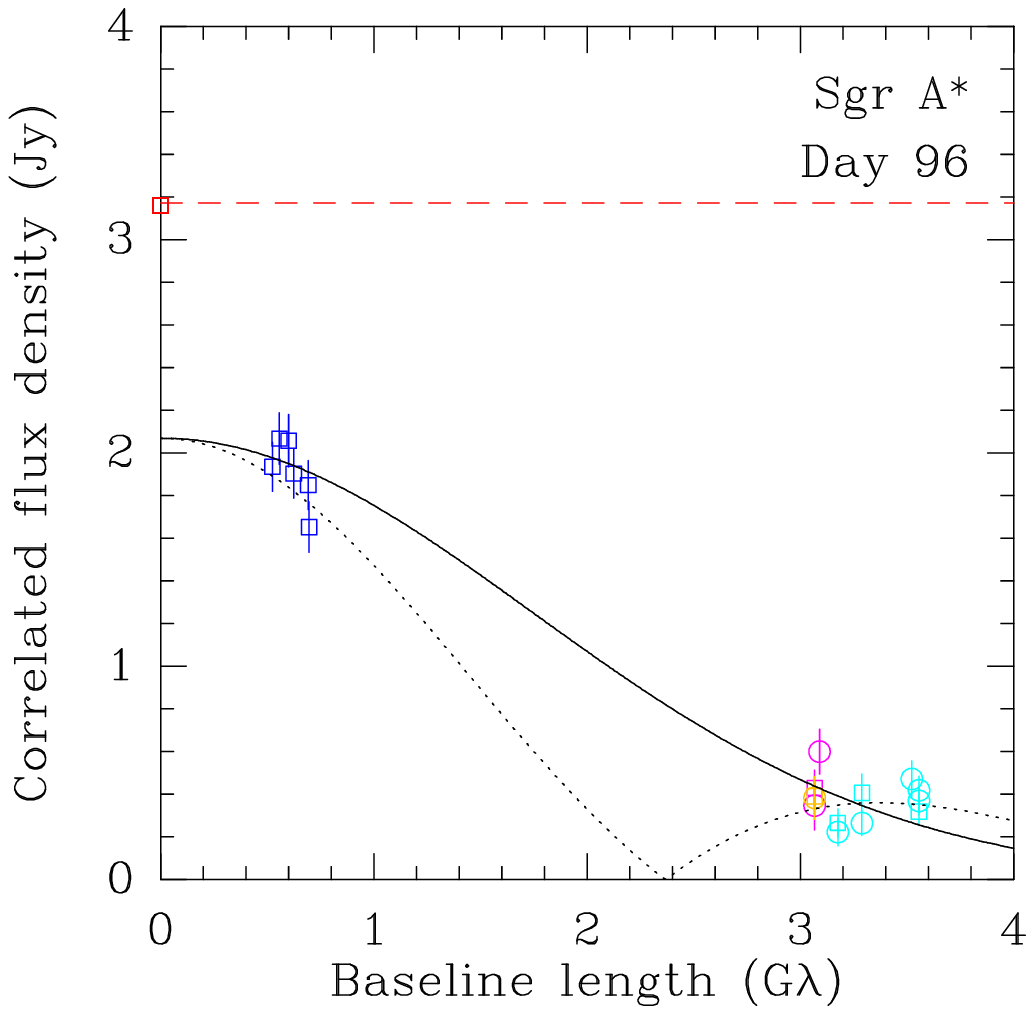}}
\resizebox{0.31\hsize}{!}{\includegraphics{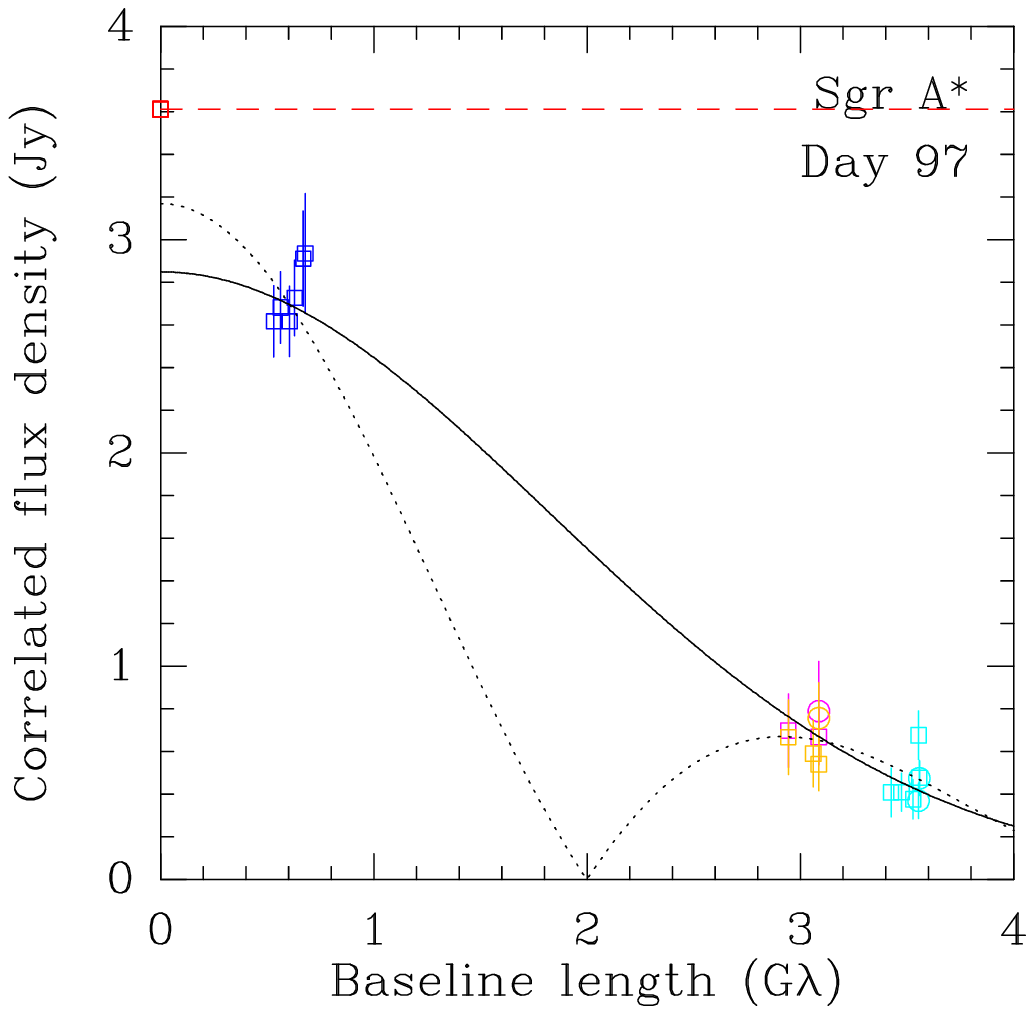}}
\caption{Correlated gain-corrected flux density plots.  Solid lines
  show best-fit circular Gaussian models of the compact emission in
  Sgr~A* and are consistent with a source size of approximately
  43~$\mu$as on all days.  The dotted lines show the best ring model
  fits to Sgr~A* data.
\label{fig-corr}
}
\end{figure*}

We report the first 1.3~mm VLBI detections of Sgr~A* on
Hawaii-California baselines with correlated flux densities for several
scans of $\gtrsim400$~mJy on the JC and JD baselines during day 96
(Figure~\ref{fig-sgra-scans}).  Nondetections on the JC and JD
baselines on day 95 are attributable to the higher opacity at the JCMT
on that day.
The robust detections on the long (Hawaii-Arizona and
Hawaii-California) baselines confirm the detection of event horizon
scale structure reported in \citet{doeleman2008}.

Because Sgr~A* was detected on baselines between all three sites, we
are able to measure closure phase: the sum of interferometric phase
around a closed triangle of baselines.  This quantity is relatively
immune to calibration errors and in general provides important
constraints on source structure.  On the CARMA-ARO/SMT-JCMT triangle,
the phase closures on Sgr A* for day 96 (5 independent measurements)
were computed using 10-second coherent integrations, which were
averaged over full 10-minute VLBI scans to increase SNR
\citep{rogers1995}.  Because of the relatively low SNR on the long
baselines to the JCMT, it is only possible to say that the closure
phases are consistent with a value of zero with a variation of
$\pm40^\circ$.

\subsection{Variability}
\label{variability}

The arcsecond-scale flux density of Sgr~A* on days 95 and 96 is the
same to within uncertainties in the flux scale, but on day 97 the flux
density of Sgr~A* was $\sim17$\% higher.  This brightening is
accompanied by changes on VLBI scales as well.  First, the SC/CD and
SD/CD flux density ratios are higher on day 97 than on days 95 and 96
(Figure~\ref{fig-sgra-scans}).  Second, there are more
Hawaii-California detections during day 97, and the apparent flux
densities on these baselines are on average higher than on day 96.
Third, the flux densities on the SJ baseline are also larger on day
97.  These differences are consistent with an episode of variability
in Sgr~A* during which the small-scale structure increased in flux
density between days 96 and 97.

\subsection{Geometrical Models of the Structure of Sgr~A*}\label{structure}

Though our data are better calibrated than in the previous epoch
\citep{doeleman2008}, the structure of Sgr~A* is poorly constrained
because millimeter VLBI detections of Sgr~A* remain limited in terms
of baseline length and orientation.  As a result, many models can be
made to fit the data: extended double sources, large rings,
combinations of large- and small-scale components, etc.

Nevertheless, with the caveat that this small dataset should not be
overinterpreted, it is instructive to investigate the two classes of
models originally considered by \citet{doeleman2008} to fit the 1.3~mm
VLBI data obtained in 2007: circular Gaussians and rings.  All of the
2007 data points could be fitted by a single Gaussian component.  In
contrast, we note a loss of $\sim1$~Jy of correlated flux density
between the connected-element (CD) and SC/SD baselines
(Figure~\ref{fig-corr}).  In the context of Gaussian models of
emission on $R_\mathrm{Sch}$ scales, this suggests the existence of
additional variable structure on scales between those probed by the
SC/SD (a few hundred microarcseconds) and the CD (a few arcseconds)
baselines.

We adopt this assumption to estimate the size of the inner accretion
flow in Sgr~A*.  Effectively, this reduces to fitting all of the VLBI
data excluding the CD data points.  For the Gaussian model, the best
fits imply a flux density of $2.07^{+0.14}_{-0.15}$~Jy and a size of
$41.3^{+5.4}_{-4.3}~\mu$as (FWHM; errors are $3\sigma$) on day 95 and
$2.07^{+0.19}_{-0.19}$~Jy and $44.4^{+3.0}_{-3.0}~\mu$as on day 96
(Figure~\ref{fig-corr} and Table~\ref{table-fits}).  These values are
consistent with the single compact component Gaussian fit of
\citet{doeleman2008}, who estimated a flux density of $2.4 \pm 0.5$~Jy
and a size of $43^{+14}_{-8}~\mu$as (before deconvolution of the
expected interstellar scattering; $37~\mu$as unscattered) for the
230~GHz emission.  On day 97, the best fit model has a much higher
flux density of $2.85^{+0.29}_{-0.28}$~Jy but a similar FWHM of
$42.6^{+3.1}_{-2.9}~\mu$as.  Despite the increase in flux density
observed on day 97, the diameter of the fitted compact component in
Sgr~A* on that day is identical (to within the errors) to the values
for the size obtained on days 95 and 96.

Ring models with three parameters (inner radius, outer radius, and
flux density) can also fit the VLBI data (Figure~\ref{fig-corr} and
Table~\ref{table-fits}).  However, no single set of ring model
parameters consistently fits the data on all three days, which would
suggest that the size and structure of Sgr~A* are variable within the
context of ring models.  This stands in contrast with the Gaussian
model, for which all epochs of data are consistent with a uniform size
despite differences in the flux density.  Longer-wavelength VLBI
observations are inconclusive as to whether a significant correlation
exists between the flux density and size of the emission in Sgr~A*
\citep{bower2004,lu2010}.  However, the size of the emission at these
wavelengths is dominated by interstellar scattering effects.

Future 1.3~mm VLBI observations with higher sensitivity, sufficient to
robustly measure the closure phase, will be an important discriminant
between these and other models.  For example, an elliptical Gaussian
distribution will result in zero closure phase on any triangle of
baselines, while a ring model can result in closure phases of 180
degrees depending on the orientation and length of the array
baselines.  The ring models shown in Figure~\ref{fig-corr} all have
closure phases of zero on the CARMA-ARO/SMT-JCMT triangle of
baselines, consistent with the measured closure phases
(Section~\ref{results}).  However, a ring model with a null near
$3.4$~$G\lambda$ between the CARMA-JCMT and ARO/SMT-JCMT baselines,
would produce a closure phase of 180 degrees, which is strongly ruled
out by the April 2009 data.  Measurement on an intermediate baseline
in the $1-2$~$G\lambda$ range would provide a powerful discriminant
between large classes of geometrical models.

\section{Discussion}

\subsection{Implications for Accretion and Flare Models}

The flux density of Sgr~A* on VLBI scales is seen to increase from day
95/96 to day 97.  During the first $\sim1$ hour on day 97, when the
atmosphere at CARMA was relatively stable, the data are consistent
with a constant flux density, suggesting that the flux density
increased before observations on day 97 but held steady at a higher
level than on the previous nights.  This behavior is consistent with
other (sub)millimeter observations, which show variability punctuated
with periods when the flux density is stable
\citep[e.g.,][]{marrone2006,yusefzadeh2009,kunneriath2010}.

The flux density increase appears to be due to an event that
establishes a new steady-state in Sgr~A*.  If instead the flux density
increase is due to a short-duration event that concluded before the
start of observations on day 97, the unchanging size of the compact
region (as implied by Gaussian models in Section~\ref{structure}) and
the timescale over which the compact flux density is seen to be
constant limits the expansion speed of the region to be highly
nonrelativistic \citep[$v \lesssim0.05 c$, consistent with][]{lu2010}
and much lower than the sound speed ($c/\sqrt{3}$)
\citep{marrone2008}, in contrast with relativistic jet models
\citep[e.g.,][]{falcke2009}.  While a low expansion speed is predicted
by adiabatically-expanding plasmon models \citep{yusefzadeh2009},
these models also predict an increase in source size.  Our
observations detect Sgr~A* after the increase in flux density has
occurred, but we do not find evidence of an increase in source size as
predicted by adiabatic expansion.  Future, more sensitive observations
of Sgr~A* before, during, and after a flare event will be necessary to
more fully understand the mechanism responsible for variability in
Sgr~A*.

\subsection{Physical Models of the Structure of Sgr~A*}\label{physical}

Numerous physically-motivated models have been advanced to explain the
multiwavelength emission from Sgr~A*
\citep[e.g.,][]{falcke2000,yuan2003,markoff2007,noble2007}.  In recent
years, several efforts to constrain disk models using 1.3~mm VLBI
observations have placed limits on model parameters such as the
inclination of the disk and the spin of the black hole.  Radiatively
inefficient accretion flow models and some general relativistic
magnetohydrodynamic simulations found that the \citet{doeleman2008}
data were already sufficient to rule out a low-inclination (i.e.,
nearly face-on) disk \citep{broderick2009,moscibrodzka2009}.
Low-inclination disk models are qualitatively similar to the ring
model in Section~\ref{structure}, with a ``null'' in the correlated
flux density as a function of baseline length whose depth decreases
with increasing disk inclination \citep{dexter2009,fish2009}.  The
location of this null as a function of baseline length is set by the
ratio of the mass of the black hole and the distance to Sgr~A*, which
is determined to within a few percent \citep{ghez2008,gillessen2009}.

Emission in high-inclination models is dominated by the highly
Doppler-boosted approaching side of the disk.  High-inclination disk
models exhibit a monotonic decrease in flux density as baselines
approach the length of the SJ baseline, with long-baseline correlated
flux densities highly dependent on the orientation of the disk in the
plane of the sky.  The decline of correlated flux density with
baseline length from Hawaii-California to Hawaii-Arizona baselines
requires that the inclination be $\gtrsim40\degr$ in several disk
models that were consistent with the 2007 data
\citep{dexter2009,broderick2009}.  The impact of the 2009
Hawaii-California detections on the 2007 fits is to strengthen the
constraint on inclination in these models (e.g., Broderick et al.,
private communication).

\section{Summary}

VLBI observations of Sgr~A* at 1.3mm wavelength in 2009 have robustly
confirmed the detection of Schwarzschild radius scale structures
reported in \citet{doeleman2008}.  On the third of three days of
observations, the total flux density of Sgr~A* was observed to have
increased by $\sim17$\%, indicating an episode of variability similar
to those described in multi-wavelength monitoring campaigns
\citep[e.g.,][]{marrone2008,yusefzadeh2009}.  The VLBI data support a
picture in which this flux density increase is due to a brightening of
structure on scales of only a few $R_\mathrm{Sch}$.  Having achieved
the dual technical milestones of successfully detecting closure phase
and developing robust methods of amplitude calibration, it is clear
that future (sub)millimeter VLBI observations with higher sensitivity
and many more baselines will be able to powerfully constrain models of
Sgr~A* on event horizon scales.

\acknowledgments

High-frequency VLBI work at MIT Haystack Observatory is supported by
grants from the National Science Foundation (NSF).  ARO receives
partial support from the NSF ATI program.  The Submillimeter Array is
a joint project between the Smithsonian Astrophysical Observatory and
the Academia Sinica Institute of Astronomy and Astrophysics and is
funded by the Smithsonian Institution and the Academia Sinica.
Funding for ongoing CARMA development and operations is supported by
the NSF and the CARMA partner universities.  DEB acknowledges support
from the NSF Research Experiences for Undergraduates program.

{\it Facilities:} \facility{CARMA ()}, \facility{HHT ()},
\facility{JCMT ()}, \facility{SMA ()}, \facility{CSO ()}

\pagebreak

\LongTables

\begin{deluxetable*}{lrrrrrrrr} 
\tabletypesize{\scriptsize} 
\tablecaption{Gain-Corrected Detections of 1924$-$292 and Sgr~A*\label{table-detections}} 
\tablehead{ 
  \colhead{} & 
  \colhead{Day} & 
  \colhead{UT Time} & 
  \colhead{} & 
  \colhead{} & 
  \colhead{$u$} & 
  \colhead{$v$} &
  \colhead{Flux Density} &
  \colhead{$\sigma$} \\
  \colhead{Source} &
  \colhead{Number} & 
  \colhead{(hh mm)} & 
  \colhead{Baseline} & 
  \colhead{Band} &
  \colhead{(M$\lambda$)} &
  \colhead{(M$\lambda$)} & 
  \colhead{(Jy)} &
  \colhead{(Jy)} 
} 
\startdata 
1924$-$292 &  95 & 13 10 & SC & both &   -509 &    105 &  4.96 &   0.26 \\
           &     &       & SJ & low  &  -3103 &  -1561 &  1.47 &   0.11 \\
           &     &       & SJ & high &  -3103 &  -1561 &  1.40 &   0.12 \\
           &     &       & JC & low  &   2594 &   1667 &  1.63 &   0.19 \\
           &     &       & JC & high &   2594 &   1667 &  1.95 &   0.22 \\
           &     &       & JD & low  &   2594 &   1667 &  1.72 &   0.20 \\
           &     &       & JD & high &   2594 &   1667 &  1.73 &   0.22 \\
           &     & 13 55 & SC & both &   -569 &    157 &  5.60 &   0.29 \\
           &     &       & SJ & low  &  -3325 &  -1251 &  1.24 &   0.09 \\
           &     &       & SJ & high &  -3325 &  -1251 &  1.13 &   0.09 \\
           &     &       & JC & low  &   2756 &   1408 &  1.33 &   0.14 \\
           &     &       & JC & high &   2756 &   1408 &  1.17 &   0.14 \\
           &     &       & JD & low  &   2756 &   1408 &  1.28 &   0.14 \\
           &     &       & JD & high &   2756 &   1408 &  1.15 &   0.14 \\
           &     & 15 25 & SC & both &   -621 &    273 &  4.30 &   0.22 \\
           &     &       & SJ & low  &  -3380 &   -598 &  1.02 &   0.07 \\
           &     &       & SJ & high &  -3380 &   -598 &  1.02 &   0.07 \\
           &     &       & JC & low  &   2759 &    871 &  2.11 &   0.13 \\
           &     &       & JC & high &   2759 &    871 &  2.16 &   0.13 \\
           &     &       & JD & low  &   2759 &    871 &  2.13 &   0.13 \\
           &     &       & JD & high &   2759 &    871 &  2.04 &   0.12 \\
           &     & 15 45 & SC & both &   -619 &    300 &  3.66 &   0.19 \\
           &     &       & SJ & low  &  -3320 &   -454 &  1.15 &   0.07 \\
           &     &       & SJ & high &  -3320 &   -454 &  1.08 &   0.07 \\
           &     &       & JC & low  &   2701 &    754 &  2.31 &   0.14 \\
           &     &       & JC & high &   2701 &    754 &  2.13 &   0.13 \\
           &     &       & JD & low  &   2701 &    754 &  2.38 &   0.15 \\
           &     &       & JD & high &   2701 &    754 &  2.23 &   0.14 \\
           &     & 16 15 & SC & both &   -608 &    339 &  3.75 &   0.20 \\
           &     &       & SJ & low  &  -3184 &   -246 &  1.23 &   0.08 \\
           &     &       & SJ & high &  -3184 &   -246 &  1.12 &   0.07 \\
           &     &       & JC & low  &   2575 &    585 &  2.56 &   0.16 \\
           &     &       & JC & high &   2575 &    585 &  2.27 &   0.15 \\
           &     &       & JD & low  &   2576 &    585 &  2.35 &   0.15 \\
           &     &       & JD & high &   2576 &    585 &  2.21 &   0.15 \\
           &  96 & 13 10 & SC & both &   -515 &    110 &  5.03 &   0.27 \\
           &     &       & SJ & low  &  -3127 &  -1535 &  1.67 &   0.12 \\
           &     &       & SJ & high &  -3127 &  -1535 &  1.64 &   0.12 \\
           &     &       & JC & low  &   2612 &   1645 &  1.96 &   0.16 \\
           &     &       & JC & high &   2612 &   1645 &  1.77 &   0.16 \\
           &     &       & JD & low  &   2612 &   1645 &  2.04 &   0.17 \\
           &     &       & JD & high &   2612 &   1645 &  1.96 &   0.17 \\
           &     & 14 00 & SC & both &   -578 &    168 &  5.78 &   0.30 \\
           &     &       & SJ & low  &  -3354 &  -1187 &  1.48 &   0.10 \\
           &     &       & SJ & high &  -3354 &  -1187 &  1.40 &   0.10 \\
           &     &       & JC & low  &   2776 &   1356 &  1.48 &   0.13 \\
           &     &       & JC & high &   2776 &   1356 &  1.38 &   0.12 \\
           &     &       & JD & low  &   2776 &   1356 &  1.47 &   0.13 \\
           &     &       & JD & high &   2776 &   1356 &  1.47 &   0.13 \\
           &     & 15 30 & SC & both &   -621 &    285 &  4.21 &   0.22 \\
           &     & 15 50 & SC & both &   -617 &    312 &  3.28 &   0.18 \\
           &     &       & SJ & low  &  -3286 &   -391 &  1.04 &   0.07 \\
           &     &       & SJ & high &  -3286 &   -391 &  0.90 &   0.07 \\
           &     &       & JC & low  &   2668 &    703 &  2.21 &   0.15 \\
           &     &       & JC & high &   2668 &    703 &  2.16 &   0.15 \\
           &     &       & JD & low  &   2668 &    703 &  2.32 &   0.15 \\
           &     &       & JD & high &   2668 &    703 &  2.19 &   0.15 \\
           &     & 16 20 & SC & both &   -603 &    351 &  3.86 &   0.25 \\
           &     &       & SJ & low  &  -3132 &   -185 &  1.09 &   0.13 \\
           &     &       & SJ & high &  -3132 &   -185 &  1.04 &   0.10 \\
           &     &       & JC & low  &   2529 &    536 &  2.54 &   0.23 \\
           &     &       & JC & high &   2529 &    536 &  2.41 &   0.21 \\
           &     &       & JD & low  &   2529 &    536 &  2.46 &   0.22 \\
           &     &       & JD & high &   2529 &    536 &  2.27 &   0.19 \\
           &  97 & 13 10 & SC & both &   -521 &    114 &  5.64 &   0.32 \\
           &     &       & SJ & low  &  -3151 &  -1509 &  1.79 &   0.14 \\
           &     &       & SJ & high &  -3151 &  -1509 &  1.53 &   0.12 \\
           &     &       & JC & low  &   2630 &   1623 &  1.79 &   0.20 \\
           &     &       & JC & high &   2630 &   1623 &  2.00 &   0.23 \\
           &     &       & JD & low  &   2630 &   1623 &  1.80 &   0.21 \\
           &     &       & JD & high &   2630 &   1623 &  1.74 &   0.21 \\
           &     & 14 00 & SC & both &   -582 &    173 &  6.09 &   0.35 \\
           &     &       & SJ & low  &  -3365 &  -1159 &  1.70 &   0.13 \\
           &     &       & SJ & high &  -3365 &  -1159 &  1.47 &   0.11 \\
           &     &       & JC & low  &   2783 &   1332 &  1.81 &   0.19 \\
           &     &       & JC & high &   2783 &   1332 &  1.97 &   0.22 \\
           &     &       & JD & low  &   2783 &   1332 &  1.83 &   0.20 \\
           &     &       & JD & high &   2783 &   1332 &  2.01 &   0.23 \\
           &     & 15 30 & SC & both &   -620 &    290 &  3.79 &   0.21 \\
           &     &       & SJ & low  &  -3344 &   -505 &  1.15 &   0.08 \\
           &     &       & SJ & high &  -3344 &   -505 &  0.95 &   0.07 \\
           &     &       & JC & low  &   2724 &    796 &  2.27 &   0.18 \\
           &     &       & JC & high &   2724 &    796 &  2.29 &   0.18 \\
           &     &       & JD & low  &   2724 &    796 &  2.39 &   0.17 \\
           &     &       & JD & high &   2724 &    796 &  2.43 &   0.18 \\
           &     & 15 50 & SC & both &   -616 &    317 &  3.53 &   0.21 \\
           &     &       & SJ & low  &  -3269 &   -364 &  1.23 &   0.09 \\
           &     &       & SJ & high &  -3269 &   -364 &  1.03 &   0.08 \\
           &     &       & JC & low  &   2653 &    681 &  2.30 &   0.20 \\
           &     &       & JC & high &   2653 &    681 &  2.24 &   0.20 \\
           &     &       & JD & low  &   2653 &    681 &  2.22 &   0.19 \\
           &     &       & JD & high &   2653 &    681 &  2.41 &   0.19 \\
           &     & 16 15 & SC & both &   -604 &    349 &  4.05 &   0.27 \\
           &     &       & SJ & low  &  -3139 &   -193 &  1.31 &   0.11 \\
           &     &       & SJ & high &  -3139 &   -193 &  1.20 &   0.11 \\
           &     &       & JC & low  &   2535 &    542 &  2.58 &   0.23 \\
           &     &       & JC & high &   2535 &    542 &  2.54 &   0.25 \\
           &     &       & JD & low  &   2535 &    542 &  2.63 &   0.23 \\
           &     &       & JD & high &   2535 &    542 &  2.48 &   0.23 \\
Sgr A*     &  95 & 11 10 & SC & both &   -474 &     85 &  2.22 &   0.13 \\
           &     & 11 30 & SC & both &   -507 &    106 &  1.86 &   0.11 \\
           &     & 12 00 & SC & both &   -550 &    140 &  1.91 &   0.11 \\
           &     & 12 20 & SC & both &   -573 &    164 &  1.85 &   0.10 \\
           &     &       & SJ & low  &  -3338 &  -1222 &  0.33 &   0.08 \\
           &     & 12 40 & SC & both &   -591 &    188 &  2.06 &   0.11 \\
           &     &       & SJ & low  &  -3391 &  -1080 &  0.36 &   0.07 \\
           &     & 13 25 & SC & both &   -617 &    246 &  2.07 &   0.11 \\
           &     &       & SJ & low  &  -3414 &   -754 &  0.32 &   0.08 \\
           &     &       & SJ & high &  -3414 &   -754 &  0.40 &   0.09 \\
           &     & 14 10 & SC & both &   -618 &    305 &  2.01 &   0.11 \\
           &     & 15 00 & SC & both &   -592 &    370 &  1.79 &   0.11 \\
           &  96 & 11 30 & SC & both &   -513 &    110 &  1.98 &   0.12 \\
           &     & 11 50 & SC & both &   -542 &    133 &  2.11 &   0.12 \\
           &     &       & SJ & high &  -3231 &  -1404 &  0.47 &   0.08 \\
           &     & 12 20 & SC & both &   -577 &    168 &  2.10 &   0.13 \\
           &     &       & SJ & high &  -3350 &  -1195 &  0.42 &   0.07 \\
           &     &       & JC & high &   2774 &   1363 &  0.61 &   0.11 \\
           &     & 12 40 & SC & both &   -595 &    193 &  1.95 &   0.12 \\
           &     &       & SJ & low  &  -3398 &  -1051 &  0.32 &   0.06 \\
           &     &       & SJ & high &  -3398 &  -1051 &  0.37 &   0.06 \\
           &     &       & JC & low  &   2804 &   1245 &  0.44 &   0.09 \\
           &     &       & JC & high &   2804 &   1245 &  0.35 &   0.12 \\
           &     &       & JD & low  &   2804 &   1245 &  0.40 &   0.09 \\
           &     &       & JD & high &   2804 &   1245 &  0.39 &   0.10 \\
           &     & 14 15 & SC & both &   -616 &    317 &  1.89 &   0.12 \\
           &     &       & SJ & low  &  -3269 &   -370 &  0.41 &   0.09 \\
           &     &       & SJ & high &  -3269 &   -370 &  0.26 &   0.06 \\
           &     & 14 35 & SC & both &   -607 &    343 &  1.69 &   0.12 \\
           &     &       & SJ & low  &  -3168 &   -233 &  0.27 &   0.07 \\
           &     &       & SJ & high &  -3168 &   -233 &  0.22 &   0.06 \\
           &  97 & 11 30 & SC & both &   -519 &    115 &  2.62 &   0.17 \\
           &     & 11 50 & SC & both &   -547 &    137 &  2.68 &   0.17 \\
           &     &       & SJ & low  &  -3250 &  -1377 &  0.38 &   0.09 \\
           &     & 12 20 & SC & both &   -581 &    173 &  2.62 &   0.16 \\
           &     &       & SJ & low  &  -3362 &  -1167 &  0.48 &   0.08 \\
           &     &       & SJ & high &  -3362 &  -1167 &  0.48 &   0.08 \\
           &     &       & JC & low  &   2781 &   1340 &  0.67 &   0.13 \\
           &     &       & JC & high &   2781 &   1340 &  0.79 &   0.23 \\
           &     &       & JD & low  &   2781 &   1340 &  0.54 &   0.12 \\
           &     &       & JD & high &   2781 &   1340 &  0.76 &   0.17 \\
           &     & 12 40 & SC & both &   -598 &    198 &  2.73 &   0.18 \\
           &     &       & SJ & low  &  -3404 &  -1023 &  0.68 &   0.11 \\
           &     &       & SJ & high &  -3404 &  -1023 &  0.37 &   0.08 \\
           &     &       & JD & low  &   2807 &   1221 &  0.59 &   0.16 \\
           &     & 13 25 & SC & both &   -619 &    257 &  2.91 &   0.22 \\
           &     &       & SJ & low  &  -3405 &   -697 &  0.41 &   0.09 \\
           &     &       & JC & low  &   2786 &    953 &  0.70 &   0.17 \\
           &     &       & JD & low  &   2786 &    953 &  0.67 &   0.17 \\
           &     & 13 40 & SC & both &   -621 &    276 &  2.94 &   0.28 \\
           &     &       & SJ & low  &  -3375 &   -589 &  0.41 &   0.11
\enddata
\end{deluxetable*}

\begin{deluxetable*}{lrrrr}
\tabletypesize{\scriptsize} 
\tablecaption{Model Fits to Sgr~A* Data\label{table-fits}} 
\tablehead{ 
  \colhead{} &
  \colhead{Day} & 
  \colhead{Compact} & 
  \colhead{Inner} &
  \colhead{Outer} \\
  \colhead{Model} &
  \colhead{Number} &
  \colhead{Flux Density} &
  \colhead{Size} &
  \colhead{Size} \\
  \colhead{} &
  \colhead{} &
  \colhead{(Jy)} &
  \colhead{($\mu$as)} &
  \colhead{($\mu$as)} 
}
\startdata
Gaussian & 95 & 2.07 & 41 & \nodata \\
         & 96 & 2.07 & 44 & \nodata \\
         & 97 & 2.85 & 43 & \nodata \\
Ring     & 95 & 2.07 & 53 &  60     \\
         & 96 & 2.07 & 37 &  92     \\
         & 97 & 3.17 & 48 & 106     
\enddata
\end{deluxetable*}


\begin{thebibliography}{}

\bibitem[Bower et al.(2004)]{bower2004} Bower, G.~C., Falcke, H.,
  Herrnstein, R.~M., Zhao, J.-H., Goss, W.~M., \& Backer, D.~C.\ 2004,
  Science, 304, 704

\bibitem[Broderick et al.(2009)]{broderick2009} Broderick, A.~E.,
  Fish, V.~L., Doeleman, S.~S., \& Loeb, A.\ 2009, \apj, 697, 45

\bibitem[Dexter et al.(2009)]{dexter2009} Dexter, J., Agol, E., \&
  Fragile, C.\ 2009, \apjl, 703, L142
  
\bibitem[Dodds-Eden et al.(2009)]{doddseden2009} Dodds-Eden, K., et
  al.\ 2009, \apj, 698, 676

\bibitem[Doeleman et al.(2001)]{doeleman2001} Doeleman, S.~S., et 
  al.\ 2001, \aj, 121, 2610
  
\bibitem[Doeleman et al.(2008)]{doeleman2008} Doeleman, S.~S., et
  al.\ 2008, \nat, 455, 78
  
\bibitem[Doeleman et al.(2009)]{doeleman2009} Doeleman, S.~S., Fish,
  V.~L., Broderick, A.~E., Loeb, A., \& Rogers, A.~E.~E.\ 2009, \apj,
  695, 59
  
\bibitem[Falcke et al.(2009)]{falcke2009} Falcke, H., Markoff, S., \&
  Bower, G.~C.\ 2009, \aap, 496, 77

\bibitem[Falcke et al.(2000)]{falcke2000} Falcke, H., Melia, F., \&
  Agol, E.\ 2000, \apjl, 528, 13

\bibitem[Fish et al.(2009a)]{fish2009} Fish, V.~L., Broderick, A.~E.,
  Doeleman, S.~S., \& Loeb, A.\ 2009a, \apjl, 692, L14

\bibitem[Fish et al.(2009b)]{fish2009b} Fish, V.~L., Doeleman, S.~S.,
  Broderick, A.~E., Loeb, A., \& Rogers, A.~E.~E.\ 2009b, \apj, 706,
  1353

\bibitem[Ghez et al.(2008)]{ghez2008} Ghez, A., et al.\ 2008, \apj,
  689, 1044

\bibitem[Gillessen et al.(2009)]{gillessen2009} Gillessen, S.,
  Eisenhauer, F., Trippe, S., Alexander, T., Genzel, R., Martins, F.,
  \& Ott, T.\ 2009, \apj, 692, 1075

\bibitem[Kunneriath et al.(2010)]{kunneriath2010} Kunneriath, D., et
  al.\ 2010, \aap, 517, A46

\bibitem[Lu et al.(2010)]{lu2010} Lu, R.~S., Krichbaum, T.~P., Eckart,
  A., K{\"o}nig, S., Kunneriath, D., Witzel, G., Witzel, A., \&
  Zensus, J.~A.\ 2010, \aap, in press, arXiv:1010.1287

\bibitem[Markoff et al.(2007)]{markoff2007} Markoff, S., Bower, G.~C.,
  \& Falcke, H.\ 2007, \mnras, 379, 1519

\bibitem[Marrone et al.(2006)]{marrone2006} Marrone, D.~P., Moran,
  J.~M., Zhao, J.-H., \& Rao, R.\ 2006, \apj, 640, 308

\bibitem[Marrone et al.(2008)]{marrone2008} Marrone, D.~P., et
  al.\ 2008, \apj, 682, 373

\bibitem[Mo{\'s}cibrodzka et al.(2009)]{moscibrodzka2009}
  Mo{\'s}cibrodzka, M., Gammie, C.~F., Dolence, J.~C., Shiokawa, H.,
  \& Leung, P.~K.\ 2009, \apj, 706, 497

\bibitem[Noble et al.(2007)]{noble2007} Noble, S.~C., Leung, P.~K.,
  Gammie, C.~F., \& Book, L.~G.\ 2007, Classical and Quantum Gravity,
  24, 259

\bibitem[Reid(2009)]{reid2009} Reid, M.\ 2009, International Journal
  of Modern Physics D, 18, 889

\bibitem[Rogers et al.(1995)]{rogers1995} Rogers, A.~E.~E., Doeleman,
  S.~S., \& Moran, J.~M.\ 1995, \aj, 109, 1391

\bibitem[Shen(1997)]{shen1997} Shen, Z.~-Q., et al.\ 1997, \aj, 114, 1999


\bibitem[Yuan et al.(2003)]{yuan2003} Yuan, F., Quataert, E., \&
  Narayan, R.\ 2003, \apj, 598, 301

\bibitem[Yusef-Zadeh et al.(2009)]{yusefzadeh2009} Yusef-Zadeh, F.,
  et al.\ 2009, \apj, 706, 348

\end{thebibliography}
\end{document}